\documentclass{book} \usepackage{amsmath} \usepackage{hhline}
\usepackage{graphicx} \usepackage[T2A]{fontenc}
\usepackage[cp1251]{inputenc} \usepackage[russian]{babel}

\begin{document} \pagestyle{headings} \parindent=7mm

\def\ds{\displaystyle} \def\ss{\scriptsize} \def\hh{\hskip 1pt}
\def\hs{\hskip 2pt} \def\h{\hskip 0.2mm} \def\pr{\prime}
\newcommand{\fbR}{\hbox{\ss\bf R}}
\newcommand{\fbr}{\hbox{\footnotesize\bf r}}
\newcommand{\fbk}{\hbox{\footnotesize\bf k}}

\phantom{x}

\setcounter{page}{1}

\makeatletter\renewcommand{\@evenhead} {\vbox{\hbox
to\textwidth{\thepage\hfil\small\it Б.В.Бондарев \hfil}\vskip
1mm\hrule\vskip -6mm}}

\renewcommand{\@evenhead}{}

\makeatletter\renewcommand{\@oddhead} {\vbox{\hbox
to\textwidth{\hfill\small\it Функция Ферми-Дирика и энергетическая щель \hfill\rm\thepage}\vskip 1mm\hrule\vskip -6mm}}

\renewcommand{\@oddhead}{}

\vskip 35mm \centerline{\large\bf FERMI – DIRAC FUNCTION AND ENERGY GAP }

\vskip 7mm \centerline{\large\bf Boris V. BONDAREV } \vskip 7mm

\centerline{\it Moscow Aviation Institute, Volokolamskoe Shosse 4, Moscow, 125871, Russia } \vskip 7mm

\centerline{E-mail: bondarev.b@mail.ru} \vskip 7mm

\begin{quote} \hskip 5mm \small Medium field method is applied for studying valence electron behavior in metals. When different wave-vector electrons are attracted at low temperatures, distribution function gets discontinued. As a result, a specific energy gap occurs. \end{quote} \vskip 7mm

\centerline{\bf 1. Introduction } \vskip 2mm

\par As it follows from the preceding paper, interacting electron distribution function has the anisotropic effect produced by the energy of interaction of electrons with opposite wave vectors $\bf k$ and $-\hs\bf k$, in particular:
$$ \varepsilon_{\bf kk^{\h\pr}}=I\hs\delta_{\hh\bf k\hh +\hh\bf k^{\h\pr}}\hh . $$
This means that at low temperatures one of the $\bf k$ and $-\hs{\bf k}$ wave vector states within particular region $S$ is more likely filled up and another one remains free. This kind of anisotropy is caused by superconductivity [1-3]. But the question is now arises: By what means does an energy gap occur? To clarify the matter we will consider another model Hamiltonian, this paper is devoted to.

\par As before, we proceed from Fermi -- Dirac function mean field approximation equation:
$$ \ln\h\frac{1-w_{\h\bf k}}{w_{\h\texttt{}\bf
k}}=\beta\h (\h\overline\varepsilon_{\bf k}-\mu)\hs , \eqno (1) $$
where $w_{\hh\fbk}$ is probability that the state described by wave function
$\psi_{\hh\fbk n}({\bf r})\hh\chi_\sigma(\xi)$
is filled by one of equilibrium system electrons with wave vector $\bf k$;
$\beta=(\hh k_{\hh\rm B}\hh T\hh)^{\h -\h 1}$, $T$ -- temperature,
$$ \overline\varepsilon_{\h\bf k}=\varepsilon_{\h\bf k} +
\sum\limits_{\bf k^\pr}\hs\varepsilon_{\h\bf kk^{\h\pr}}\hs
w_{\hh\bf k^{\h\pr}} \eqno (2) $$
-- electron medium energy in one of the Bloch states with wave vector $\bf k$,
$\varepsilon_{\h\bf k}$ is electron energy less interaction with other electrons, $\varepsilon_{\h\bf kk^{\h\pr}}$ is energy of interaction of electrons with wave vectors $\bf k$ and $\bf k^{\h\pr}$, $\mu$ is chemical potential.

\vskip 4mm \centerline{\bf 2. Model Hamiltonian } \vskip 2mm

\par We consider the following Hamiltonian [4, 5]
$$ \varepsilon_{\bf kk^{\h\pr}}=\hh -\hs J\hs\delta_{\hh\bf k\hh -\hh\bf k^{\h\pr}}
\hh . \eqno (3) $$
Here, $\delta_{\hh\bf k\hh -\hh\bf k^{\h\pr}}$ is a Kronecker delta, $J$ -- electron attraction energy with different wave vectors $\bf k$ and $\bf k^{\h\pr}={\bf k}$. As provided by this model, those valence electrons moving within their crystal space at equal speeds may be attracted only. Such Hamiltonian is produced by Coulomb electron attraction.

\par We use electron interaction energy (3) in formula (2) and get the equation as follows:
$$ \overline\varepsilon_{\h\bf k}=\varepsilon_{\h\bf k} -
J\hs w_{\hh\bf k}\hh . \eqno (4) $$
Now, equation (1) takes on the form as follows:
$$ \theta\hs\ln\h\frac{1-w_{\h\bf k}}{w_{\h\texttt{}\bf
k}}=\varepsilon_{\h\bf k} - J\hs w_{\hh\bf k} -\mu\hh , \eqno (5) $$
where $\theta=k_{\hh\rm B}\hh T$.

\vskip 4mm \centerline{\bf 3. Distribution of Attracting Electrons } \vskip 2mm

\par Equation (5) may be solved by the form as follows:
$$ w_{\hh\bf k}=w(x_{\h\bf k},\hh\tau)\hh , \eqno (6) $$
where
$$ x_{\h\bf k}=\frac{\hh\varepsilon_{\hh\bf k}-\mu\hh}{J}\hh , \eqno (7) $$
$$ \tau =\frac{\hh 4\hs\theta\hh}{J}\hh . \eqno (8) $$
Wherein, relation $w=w(x,\hh\tau)$ is given by the following equation:
$$ x=w+\frac{\hh\tau\hh}{4}\hs\ln\hh\frac{\hs 1-w\hs}{w}\hh . \eqno (9) $$
For relation diagrams of $w=w(x,\hh\tau)$ regarding various temperature values $\tau$, see Fig. 1.

\makeatletter\renewcommand{\@oddhead} {\vbox{\hbox
to\textwidth{\thepage\hfil\small\it B.V.Bondarev \hfil}\vskip
1mm\hrule\vskip -3mm}}

\makeatletter\renewcommand{\@evenhead} {\vbox{\hbox
to\textwidth{\hfill\small\it Fermi – Dirac Function and Energy Gap \hfill\rm\thepage}\vskip 1mm\hrule\vskip -3mm}}

\vskip 10mm \unitlength=1mm \centerline{\begin{picture}(85,57)
\put(24,17){\it 1}\put(32,17){\it 2}\put(43,17){\it
3}\put(61,17){\it 4} \put(0,10){\vector(1,0){85}}
\put(65.5,5){$x$\hs=\hs$(\varepsilon$\hs--\hs$\mu)/J$}
\put(20,10){\line(0,-1){1}}\put(19.2,5){0}
\put(40,10){\line(0,-1){1}}\put(37.7,5){0,5}
\put(60,10){\line(0,-1){1}}\put(59.2,5){1}
\put(20,10){\vector(0,1){50}}\put(22,58){$w(x,\hh\tau)$}
\put(14,11){$0$} \put(20,30){\line(1,0){1}}\put(14,29){$0,5$}
\put(14,51.5){$1,0$}
\multiput(60,10.5)(0,2.013){20}{\line(0,1){1.4}}
\put(20,10){\unitlength=1mm\special{em:linewidth 0.3pt}
\put(-20,40){\special{em:moveto}} \put(40,40) {\special{em:lineto}}
\put(0,0)   {\special{em:lineto}} }
\put(20,10){\unitlength=1mm\special{em:linewidth 0.3pt}
\put(15.27,39.7){\special{em:moveto}}
\put(17.65,39.5){\special{em:lineto}}
\put(19.16,39.3){\special{em:lineto}} \put(20.68,39)
{\special{em:lineto}} \put(22.01,38.6){\special{em:lineto}}
\put(23.28,38)  {\special{em:lineto}} \put(24.44,37)
{\special{em:lineto}} \put(25.02,36)  {\special{em:lineto}}
\put(25.27,35)  {\special{em:lineto}} \put(25.33,34)
{\special{em:lineto}} \put(25.24,33)  {\special{em:lineto}}
\put(25.07,32)  {\special{em:lineto}} \put(24.82,31)
{\special{em:lineto}} \put(24.50,30)  {\special{em:lineto}}
\put(23.76,28)  {\special{em:lineto}} \put(22.90,26)
{\special{em:lineto}} \put(21.97,24)  {\special{em:lineto}}
\put(21.00,22)  {\special{em:lineto}} \put(20.00,20)
{\special{em:lineto}} \put(19.00,18)  {\special{em:lineto}}
\put(18.03,16)  {\special{em:lineto}} \put(17.10,14)
{\special{em:lineto}} \put(16.24,12)  {\special{em:lineto}}
\put(15.50,10)  {\special{em:lineto}} \put(15.18,9)
{\special{em:lineto}} \put(14.93,8)   {\special{em:lineto}}
\put(14.76,7)   {\special{em:lineto}} \put(14.68,6)
{\special{em:lineto}} \put(14.73,5)   {\special{em:lineto}}
\put(14.98,4)   {\special{em:lineto}} \put(15.56,3)
{\special{em:lineto}} \put(16.72,2)   {\special{em:lineto}}
\put(17.86,1.45){\special{em:lineto}} \put(19.32,1.0)
{\special{em:lineto}} \put(20.84,0.7) {\special{em:lineto}}
\put(22.35,0.5) {\special{em:lineto}} }
\put(20,10){\unitlength=1mm\special{em:linewidth 0.3pt}
\put(-4.18,39.5)  {\special{em:moveto}} \put(2.36,39)
{\special{em:lineto}} \put(8.56,38)  {\special{em:lineto}}
\put(11.88,37)  {\special{em:lineto}} \put(14.03,36)
{\special{em:lineto}} \put(15.54,35)  {\special{em:lineto}}
\put(16.66,34)  {\special{em:lineto}} \put(18.14,32)
{\special{em:lineto}} \put(19.01,30)  {\special{em:lineto}}
\put(19.53,28)  {\special{em:lineto}} \put(19.81,26)
{\special{em:lineto}} \put(19.94,24)  {\special{em:lineto}}
\put(20.00,22)  {\special{em:lineto}} \put(20.00,20)
{\special{em:lineto}} \put(20.00,18)  {\special{em:lineto}}
\put(20.06,16)  {\special{em:lineto}} \put(20.19,14)
{\special{em:lineto}} \put(20.47,12)  {\special{em:lineto}}
\put(20.99,10)  {\special{em:lineto}} \put(21.60,8.5)
{\special{em:lineto}} \put(21.86,8.0)  {\special{em:lineto}}
\put(22.16,7.5)  {\special{em:lineto}} \put(22.51,7.0)
{\special{em:lineto}} \put(22.90,6.5)  {\special{em:lineto}}
\put(23.35,6.0)  {\special{em:lineto}} \put(23.86,5.5)
{\special{em:lineto}} \put(24.46,5.0)  {\special{em:lineto}}
\put(25.15,4.5)  {\special{em:lineto}} \put(25.97,4.0)
{\special{em:lineto}} \put(26.95,3.5)  {\special{em:lineto}}
\put(28.12,3.0)  {\special{em:lineto}} \put(29.58,2.5)
{\special{em:lineto}} \put(31.44,2.0)  {\special{em:lineto}}
\put(33.95,1.5)  {\special{em:lineto}} \put(37.64,1.0)
{\special{em:lineto}} \put(44.20,0.5)  {\special{em:lineto}} }
 \put(20,10){\unitlength=1mm\special{em:linewidth
0.3pt} \put(-20.00,37.9)  {\special{em:moveto}} \put(-16.66,37.5)
{\special{em:lineto}} \put(-13.24,37)  {\special{em:lineto}}
\put(-10.42,36.5)  {\special{em:lineto}} \put(-7.96,36)
{\special{em:lineto}} \put(-5.82,35.5)  {\special{em:lineto}}
\put(-3.92,35)  {\special{em:lineto}} \put(-2.22,34.5)
{\special{em:lineto}} \put(-0.68,34)  {\special{em:lineto}}
\put(0.70,33.5)  {\special{em:lineto}} \put(1.98,33)
{\special{em:lineto}} \put(3.18,32.5)  {\special{em:lineto}}
\put(4.28,32)  {\special{em:lineto}} \put(8.02,30)
{\special{em:lineto}} \put(11.06,28)  {\special{em:lineto}}
\put(13.62,26)  {\special{em:lineto}} \put(15.89,24)
{\special{em:lineto}} \put(17.99,22)  {\special{em:lineto}}
\put(20.00,20)  {\special{em:lineto}} \put(22.01,18)
{\special{em:lineto}} \put(24.11,16)  {\special{em:lineto}}
\put(26.38,14)  {\special{em:lineto}} \put(28.94,12)
{\special{em:lineto}} \put(31.98,10)  {\special{em:lineto}}
\put(35.72,8)  {\special{em:lineto}} \put(38.02,7)
{\special{em:lineto}} \put(39.30,6.5)  {\special{em:lineto}}
\put(40.72,6)  {\special{em:lineto}} \put(42.24,5.5)
{\special{em:lineto}} \put(43.92,5)  {\special{em:lineto}}
\put(45.84,4.5)  {\special{em:lineto}} \put(47.96,4)
{\special{em:lineto}} \put(50.44,3.5)  {\special{em:lineto}}
\put(53.24,3)  {\special{em:lineto}} \put(56.68,2.5)
{\special{em:lineto}} \put(60.88,2)  {\special{em:lineto}} }
\end{picture}}

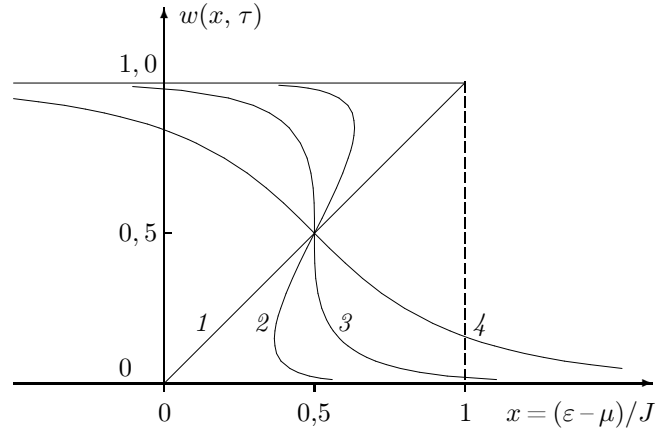
\vskip -2mm\centerline {\it Fig. 1. Electron energy distribution function at various temperature values $\tau$: }
\centerline {\it 1 -- $\tau=0$; 2 -- $\tau=0.5$; 3 -- $\tau=1$; 4 -- $\tau=2$. }
\vskip 4mm

\par At temperatures $T$ above critical point:
$$ T_c=\frac{J}{\hh 4\hs k_{\hh\rm B}} \eqno (10) $$
($\tau\geq 1$), distribution function $w=w(\varepsilon,\hh\tau)$ has a monotonic decreasing quantity. Should at a particular range $(\varepsilon_1,\hh\varepsilon_2)$ of kinetic electron energies $T<T_c$, the distribution function gets three-valued. Naturally, probability $w(\varepsilon)$ of filling with electron may gain one of the three possible values only. Actually, electrons are distributed by states provided that electron energy takes on the least value.

\vskip 4mm \centerline{\bf 4. Real Electron Distribution } \vskip 2mm

\par Now, we consider different electron equilibrium states at $T=0$. One of the distribution functions, which describe the state of electron system at $T=0$ takes the form as follows:
$$ w(\varepsilon,\hh 0)=\left\{\begin{array}{rcl} 1 & \hbox{at} &
\varepsilon<\mu+J\hh , \\ 0 & \hbox{at} &
\varepsilon>\mu\hh . \\ \end{array}\right. \eqno (11) $$
Other electron system distribution function gets the state described by the following equation:
$$ w(\varepsilon,\hh 0)=\left\{\begin{array}{ccl} 1 & \hbox{at} &
\varepsilon<\mu\hs , \smallskip \\ \ds\frac{\hh\varepsilon
-\mu\hh}{J} & \hbox{at} & \mu<\varepsilon<\mu +J\hs , \smallskip \\
0 & \hbox{at} & \varepsilon>\mu+J\hs . \\ \end{array}\right. \eqno (12) $$

\par Energy $E_{\hh 0}$ of electrons in state (11) is less than energy $E_{\h 1}$ of electrons in state (12) by the following quantity:
$$ E_{\h 1}-E_{\hh 0}=\frac{7\hs N\hs J^{\h 2}}{\hh 16\hs\varepsilon_{\hh\rm F}\hh}
\hh , \eqno (13) $$
where $N$ is number of electrons, $\varepsilon_{\hh\rm F}$ -- Fermi energy. Thus, state (11) is a basic one of electron system, i.e. the least energy state.

\par At temperatures $T<T_c$ state of the electron system which matches the least energy $E$ is described by the distribution function shown in Fig. 2. This function gets nonremovable discontinuity at the kinetic energy of electron as follows:
$$ \varepsilon =\mu+\frac{\hh J\hh}{2}\hh . \eqno (14) $$

\vskip 7mm \unitlength=1mm \centerline{\begin{picture}(65,57)
\put(0,10){\vector(1,0){63}}
\put(56.3,5){$\varepsilon$\hs--\hs$\mu$}
\put(10,10){\line(0,-1){1}}\put(9.2,5){0}
\put(30,10){\line(0,-1){1}}\put(27,5){$\frac{\hh 1\hh}{2}\hh J$}
\put(10,10){\vector(0,1){50}}\put(12,58){$w(\varepsilon,\hh\tau)$}
\put(7,11){$0$}\put(7,51.5){1}
\multiput(30,10.5)(0,2.01){20}{\line(0,1){1.4}}
\put(30,18){\circle*{0.7}}\put(30,42){\circle*{0.7}}
\multiput(30,18)(2.03,0){3}{\line(1,0){1.4}}
\multiput(30,42)(2.03,0){3}{\line(1,0){1.4}}
\put(34.9,18){\vector(0,1){24}}\put(36,29){$\Delta w$}
\put(34.9,22){\vector(0,-1){4}}
\put(10,10){\unitlength=1mm\special{em:linewidth 0.3pt}
\put(20,8.0){\special{em:moveto}}
\put(20.18,7.0){\special{em:lineto}}
\put(20.75,6.0){\special{em:lineto}}
\put(21.54,5.0){\special{em:lineto}}
\put(22.68,4.0){\special{em:lineto}}
\put(24.35,3.0){\special{em:lineto}}
\put(25.52,2.5){\special{em:lineto}}
\put(26.22,2.25){\special{em:lineto}}
\put(27.02,2.0){\special{em:lineto}}
\put(27.98,1.75){\special{em:lineto}}
\put(29.08,1.50){\special{em:lineto}}
\put(30.44,1.25){\special{em:lineto}}
\put(32.14,1.00){\special{em:lineto}}
\put(34.40,0.75){\special{em:lineto}}
\put(37.64,0.50){\special{em:lineto}}
\put(43.32,0.25){\special{em:lineto}} }
\multiput(0,50)(2.03,0){15}{\line(1,0){1.4}}
\put(9.8,10){\unitlength=1mm\special{em:linewidth 0.3pt} \put(20,32)
{\special{em:moveto}} \put(20.22,32) {\special{em:lineto}}
\put(19.82,33)  {\special{em:lineto}} \put(19.26,34)
{\special{em:lineto}} \put(18.46,35) {\special{em:lineto}}
\put(17.32,36)  {\special{em:lineto}} \put(15.65,37)
{\special{em:lineto}} \put(14.48,37.50) {\special{em:lineto}}
\put(13.78,37.75)  {\special{em:lineto}} \put(12.98,38.00)
{\special{em:lineto}} \put(12.03,38.25) {\special{em:lineto}}
\put(10.92,38.50)  {\special{em:lineto}} \put(9.564,38.75)
{\special{em:lineto}} \put(7.856,39.00) {\special{em:lineto}}
\put(5.608,39.25)  {\special{em:lineto}} \put(2.364,39.50)
{\special{em:lineto}} \put(-8.0,39.70) {\special{em:lineto}} }
\end{picture}}

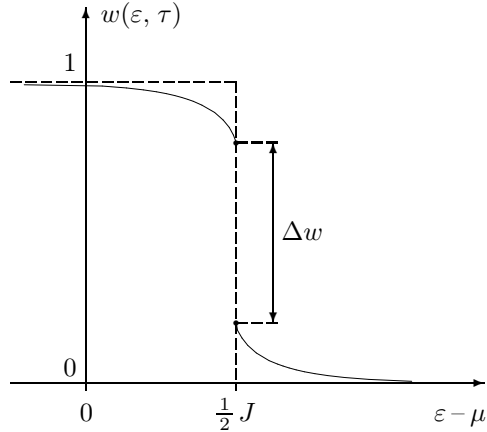
\vskip -3mm\centerline {\it Fig. 2. Real electron energy distribution function at $\tau=0.85$ } \vskip 3mm

\vskip 4mm \centerline{\bf 5. Energy Gap } \vskip 2mm

\par Gap width $\Delta w$ of distribution function may be found by means of equation (9) using $x=1/2$. On making simple calculations we obtain the equation as follows:
$$ \ln\hh\frac{\hs 1+\Delta w\hs}{\hs 1-\Delta
w\hs}= \frac{\hs 2\hs\Delta w\hs}{\tau}\hs , \eqno (15) $$
which describes relation of gap width $\Delta w$ to temperature. For the relation diagram see Fig. 3.

\newpage

\vskip 4mm \centerline{\bf 6. Medium Electron Energy } \vskip 2mm

\par Medium electron energy is determined by formula (4). As provided by this formula, electron medium energy $\overline\varepsilon_{\hh\fbk}$, like distribution function $w_{\hh\fbk}$, is a wave $\bf k$ vector composed function, where kinetic electron energy $\varepsilon_{\hh\fbk}$ is applied as an intervening variable:
$$ \overline\varepsilon(\varepsilon)=\varepsilon-J\hs w(\varepsilon)
\hh . \eqno (16) $$
For the relation diagram at $\tau =0,85$, see Fig. 4.

\phantom{}
\vskip 5mm\unitlength=1mm \centerline{\begin{picture}(74,57)
\put(12,10){\vector(1,0){59.5}}\put(69.5,5){$\tau$}
\put(12,10){\line(0,-1){1}}\put(11.2,5){0}
\put(32,10){\line(0,-1){1}}\put(29.7,5){0,5}
\put(52,10){\line(0,-1){1}}\put(51.2,5){1}
\put(12,10){\vector(0,1){50}}\put(4,58){$\Delta w$}
\put(12,10){\line(-1,0){1}}\put(4,9){$0$}
\put(12,30){\line(-1,0){1}}\put(4,29){$0,5$}
\put(12,50){\line(-1,0){1}}\put(4,49){$1,0$}
\put(12,10){\unitlength=1mm\special{em:linewidth 0.3pt}
\put(40.000,0)  {\special{em:moveto}} \put(39.967,2)
{\special{em:lineto}} \put(39.886,4)  {\special{em:lineto}}
\put(39.698,6)  {\special{em:lineto}} \put(39.461,8)
{\special{em:lineto}} \put(39.152,10)  {\special{em:lineto}}
\put(38.770,12)  {\special{em:lineto}} \put(38.310,14)
{\special{em:lineto}} \put(37.767,16)  {\special{em:lineto}}
\put(37.136,18)  {\special{em:lineto}} \put(36.410,20)
{\special{em:lineto}} \put(35.577,22)  {\special{em:lineto}}
\put(34.625,24)  {\special{em:lineto}} \put(33.536,26)
{\special{em:lineto}} \put(32.284,28)  {\special{em:lineto}}
\put(30.834,30)  {\special{em:lineto}} \put(29.128,32)
{\special{em:lineto}} \put(27.067,34)  {\special{em:lineto}}
\put(24.453,36)  {\special{em:lineto}} \put(23.830,36.40)
{\special{em:lineto}} \put(23.159,36.80)  {\special{em:lineto}}
\put(22.431,37.20)  {\special{em:lineto}} \put(21.634,37.60)
{\special{em:lineto}} \put(20.745,38)  {\special{em:lineto}}
\put(19.734,38.40)  {\special{em:lineto}} \put(18.544,38.80)
{\special{em:lineto}} \put(17.062,39.20)  {\special{em:lineto}}
\put(16.76,39.20)  {\special{em:lineto}} \put(14.80,39.60)
{\special{em:lineto}} \put(13.20,39.80)  {\special{em:lineto}}
\put(10.50,39.96)  {\special{em:lineto}} \put(0,40)
{\special{em:lineto}} } \end{picture}}

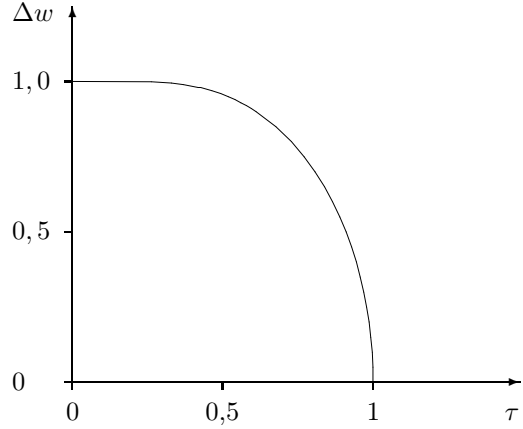
\vskip -2mm\centerline {\it Fig. 3. Relation of distribution function gap width
$\Delta w$ to temperature $\tau$ } \vskip 2mm

\vskip 10mm\unitlength=1mm \centerline{\begin{picture}(57,77)
\put(0,50){\vector(1,0){59}}\put(52.3,46.7){$\varepsilon$\hs--\hs$\mu$}
\put(7,46){0}\put(26.5,46){$\frac{J}{\hh 2\hh}$}
\put(10,7){\vector(0,1){75}}\put(12,80){$\overline\varepsilon$\hs--\hs$\mu$}
\put(10,10){\line(-1,0){1}}\put(3,9){$-\hh J$}
\multiput(10,10)(1.4,1.4){24}{\unitlength=1mm\special{em:linewidth
0.3pt} \put(0,0){\special{em:moveto}} \put(1,1){\special{em:lineto}}
} \multiput(10,50)(1.4,1.4){23}{\unitlength=1mm\special{em:linewidth
0.3pt} \put(0,0){\special{em:moveto}} \put(1,1){\special{em:lineto}}
} \multiput(30,50.4)(0,2.1){6}{\line(0,1){1.4}}
\multiput(30,49.6)(0,-2.1){6}{\line(0,-1){1.4}}
\put(30,37.2){\circle*{0.7}}\put(30,62.7){\circle*{0.7}}
\multiput(10,37.2)(2,0){10}{\line(1,0){1.4}}\put(3,36){$-\hh\Delta$}
\multiput(10,62.7)(2,0){10}{\line(1,0){1.4}}\put(6,61.5){$\Delta$}
\put(10,50){\unitlength=1mm\special{em:linewidth 0.3pt}
\put(20.00,-13.00){\special{em:moveto}}
\put(19.82,-13.18){\special{em:lineto}}
\put(19.26,-14.74){\special{em:lineto}}
\put(18.46,-16.54){\special{em:lineto}}
\put(17.32,-18.68){\special{em:lineto}}
\put(15.65,-21.35){\special{em:lineto}}
\put(14.48,-23.02){\special{em:lineto}}
\put(13.78,-23.97){\special{em:lineto}}
\put(12.98,-25.02){\special{em:lineto}}
\put(12.03,-26.22){\special{em:lineto}}
\put(10.92,-27.58){\special{em:lineto}}
\put(9.564,-29.19){\special{em:lineto}}
\put(7.856,-31.14){\special{em:lineto}}
\put(5.608,-33.64){\special{em:lineto}}
\put(2.364,-37.14){\special{em:lineto}} }
\put(10,50){\unitlength=1mm\special{em:linewidth 0.3pt}
\put(20.00,13.10){\special{em:moveto}}
\put(20.18,13.18){\special{em:lineto}}
\put(20.75,14.75){\special{em:lineto}}
\put(21.54,16.54){\special{em:lineto}}
\put(22.68,18.68){\special{em:lineto}}
\put(24.35,21.35){\special{em:lineto}}
\put(25.52,23.02){\special{em:lineto}}
\put(26.22,23.97){\special{em:lineto}}
\put(27.02,25.02){\special{em:lineto}}
\put(27.98,26.22){\special{em:lineto}}
\put(29.08,27.58){\special{em:lineto}}
\put(30.44,29.19){\special{em:lineto}}
\put(32.14,31.14){\special{em:lineto}}
} \end{picture}} \vskip -2mm

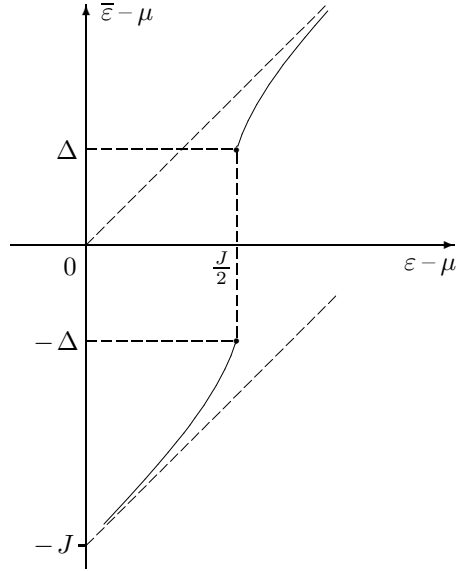
\centerline {\it Fig. 4. Relation of medium electron energy $\overline\varepsilon$ to its kinetic energy $\varepsilon$ }
\centerline {\it at temperature $\tau=0.85$ }\vskip 3mm

\par As it seen from Fig. 4, electron energy spectrum has a “discontinuity” which width $2\hh\Delta$ depends on gap width $\Delta w$ of the distribution function and determined by the ratio as follows:
$$ 2\hh\Delta =J\hs\Delta w\hs . \eqno (17) $$
At temperature $T=0$ the energy gap acquires the largest width. Should the temperature go up, gap width shall monotonically decrease. At $T=T_c$, the gap shall disappear, since distribution function $w=w(\varepsilon,\hh\tau)$ gets continuous.

\vskip 4mm \centerline{\bf 7. Conclusion } \vskip 2mm

\par We hereby have stated that model Hamiltonian (3) together with equation (1) explicitly results in occurrence of an energy gap. This model Hamiltonian is applied for stipulating the state proving availability of a gap only. The real Hamiltonian takes the form as follows [4, 5]
$$ \varepsilon_{\bf kk^{\h\pr}}=I\hs\delta_{\hh\bf k\hh +\hh\bf k^{\h\pr}}
-\hs J\hs\delta_{\hh\bf k\hh -\hh\bf k^{\h\pr}}\hh . $$
As provided by the preceding paper, the first term of the formula defines anisotropy and results in superconductivity. As for the second term, it results in occurrence of a gap.

\vskip 5mm\centerline {\bf References } \vskip 2mm

\noindent [1] H. Kamerlingh Onnes, Comm. Phys. Leb. Univ. Leiden., 1911, № 122, p. 13.

\noindent [2] J. Bardeen, L.N. Cooper, J.R. Schrieffer, Phys. Rev., 1957, v. 106, № 1,
p. 162; 1957, v. 108, № 5, p. 1175.

\noindent [3] J. Schrieffer, Superconductivity theory, M.: Science, 1970.

\noindent [4] B.V. Bondarev, Vestnik of MAI, 1996, v. 3, № 2, v. 56-65.

\noindent [5] B.V. Bondarev, Density matrix method in cooperative phenomena quantum theory, M.: Sputnik$^+$, 2001, \linebreak 250 p.

\end{document}